\documentclass{article} %
\usepackage{etoolbox}       %

\usepackage{amsmath,amsthm}   %
\usepackage{mathtools}  %
\usepackage[dvipsnames]{xcolor}         %
\usepackage[utf8]{inputenc} 
\usepackage[T1]{fontenc}    

\newtoggle{arxiv}
\toggletrue{arxiv}
\iftoggle{arxiv}{
  \usepackage[parfill]{parskip}
  \usepackage[authoryear]{natbib}

  \setlength{\textwidth}{6.8in}  %
  \setlength{\textheight}{9in}
  \setlength{\oddsidemargin}{0in}
  \setlength{\evensidemargin}{0in}
  \setlength{\topmargin}{-0.5in}
  \newlength{\defbaselineskip}
  \setlength{\defbaselineskip}{\baselineskip}
  \setlength{\marginparwidth}{0.8in}
  \setlength{\parskip}{6pt}%
  \setlength{\parindent}{0pt}%

  \RequirePackage[T1]{fontenc}
  \RequirePackage[tt=false, type1=true]{libertine}
  \RequirePackage[varqu]{zi4}
  \RequirePackage[libertine]{newtxmath}

}{
  \usepackage{amsfonts}       %
\usepackage{colm2024_conference}
  \addtolength{\tabcolsep}{-3pt} %
  \def\setstretch#1{\renewcommand{\baselinestretch}{#1}}
  \setstretch{0.98}
  \addtolength{\parskip}{-1pt}

  \usepackage[compact]{titlesec}
  \titlespacing{\section}{0pt}{*1}{*0}
  \titlespacing{\subsection}{0pt}{*1}{*0}
  \usepackage[subtle, mathdisplays=tight, charwidths=tight, leading=normal]{savetrees}
  \addtolength\textfloatsep{-0.5em}
  \addtolength\intextsep{-0.2em}
\bibliographystyle{colm2024_conference}
}

\usepackage{booktabs}       
\usepackage{amsfonts}       
\usepackage{nicefrac}       
\usepackage{microtype}      

\usepackage{multirow}
\usepackage{graphicx}
\usepackage{subcaption}
\usepackage{xcolor,colortbl}
\definecolor{Gray}{gray}{0.89}
\definecolor{LightCyan}{rgb}{0.88,1,1}
\usepackage{tabularx}
\usepackage{listings}
\usepackage{fontawesome5} 
\usepackage{makecell}
\usepackage{wrapfig}
\usepackage{amsmath,amsthm}
\usepackage{mathtools, nccmath}
\usepackage{algorithm}
\usepackage{algpseudocode}
\usepackage{tablefootnote}

\usepackage{threeparttable}
\usepackage{xspace}  

\DeclarePairedDelimiterX{\norm}[1]{\lVert}{\rVert}{#1}
\newcommand{\ie}{\emph{i.e.,}\xspace}
\newcommand{\eg}{\emph{e.g.,}\xspace}
\definecolor{mypink}{RGB}{220,20,120}
\definecolor{mygreen}{RGB}{0,130,0}
\definecolor{myblue}{RGB}{0,160,230}

\usepackage{url}
\usepackage{hyperref}
\hypersetup{
     colorlinks   = true,
     linkcolor    = blue,
     citecolor    = magenta,
     urlcolor     = red,
     pdfborderstyle={/S/U/W 1},
}

\usepackage[capitalise,noabbrev]{cleveref}  %

\usepackage{pifont}%

\iftoggle{arxiv}{
  \title{ELANA: A Simple \textbf{\underline{E}}nergy and \textbf{\underline{L}}atency \textbf{\underline{Ana}}lyzer for LLMs}
  \usepackage{authblk}
  \author[]{Hung-Yueh Chiang}
  \author[]{Bokun Wang}
  \author[]{Diana Marculescu}
  \affil[]{
    Chandra Family Department of Electrical and Computer Engineering, \protect\\ 
    The University of Texas at Austin
  }
  \affil[]{
      {\texttt{\{hungyueh.chiang, bokun.wang, dianam\}@utexas.edu}}
    }
  \date{}
}{
  \title{ELANA: A Simple \textbf{\underline{E}}nergy and \textbf{\underline{L}}atency \textbf{\underline{Ana}}lyzer for LLMs}
}

\begin{document}

\maketitle
\begin{abstract}
\noindent
The latency and power consumption of large language models (LLMs) are major constraints when serving them across a wide spectrum of hardware platforms, from mobile edge devices to cloud GPU clusters.
Benchmarking is crucial for optimizing efficiency in both model deployment and next-generation model development.
To address this need, we open-source a simple profiling tool, \textbf{ELANA}, for evaluating LLMs.
ELANA is designed as a lightweight, academic-friendly profiler for analyzing model size, key-value (KV) cache size, prefilling latency (Time-to-first-token, TTFT), generation latency (Time-per-output-token, TPOT), and end-to-end latency (Time-to-last-token, TTLT) of LLMs on both multi-GPU and edge GPU platforms.
It supports all publicly available models on Hugging Face and offers a simple command-line interface, along with optional energy consumption logging.
Moreover, ELANA is fully compatible with popular Hugging Face APIs and can be easily customized or adapted to compressed or low bit-width models, making it ideal for research on efficient LLMs or for small-scale proof-of-concept studies.
We release the ELANA profiling tool at: \url{https://github.com/enyac-group/Elana}.
\end{abstract}

\section{Introduction}
Numerous emerging applications are powered by large language models (LLMs) today.
Yet, serving models with parameters on the order of billions (\eg 100B) poses significant challenges in meeting the required inference latency, memory, and energy costs.
Extensive research has investigated quantization \citep{xiao2023smoothquant, lin2024awq, lin2024qserve, chiang2025quamba2} and compression \citep{wang2025svdllm, lin2025modegpt, chiang2025uniql} techniques to reduce inference latency and model size for deployment.
However, these research directions primarily focus on algorithmic design while overlooking energy consumption in their evaluations.
Furthermore, existing profiling benchmarks and results depend heavily on tools developed individually by researchers, and a unified and fair profiling framework is still lacking.
%

Recent work has studied the energy footprint of machine learning systems, including measuring~\citep{you2023zeus,tschand2025mlperf}, benchmarking~\citep{chung2025ml,samsi2023words,krupp2025promoting}, and optimizing~\citep{you2023zeus,chung2024reducing} energy costs for training and inference, across cloud servers and edge devices.
%
In particular, Zeus~\citep{you2023zeus} provides a general-purpose GPU energy/time profiler for any Python code block.  It also includes a command-line interface (CLI) that reports the total energy consumed by the GPU during the lifetime of the monitor process. 
%
Nevertheless, a lightweight CLI profiler tailored for standardized LLM inference that combines fine-grained latency, energy measurement, and kernel-level analysis for machine learning developers remains missing.
%

To fill this gap, we release \textbf{ELANA}, a streamlined profiling framework for benchmarking LLMs.
ELANA provides an academic-friendly interface for measuring key performance metrics for evaluating LLMs, including model size, KV cache footprint, prefilling latency (Time-to-First-Token, TTFT), generation latency (Time-per-Output-Token, TPOT), and end-to-end inference latency of requests (Time-to-Last-Token, TTLT) across both multi-GPU and edge GPU devices.
Our tool provide the features for profiling energy costs, such as Joule-per-token (J/Token), Joule-per-prompt (J/Prompt), and Joule-per-request (J/Request).
The tool supports all models available on Hugging Face and includes a minimal command-line interface.
In addition, ELANA integrates seamlessly with Hugging Face APIs and can be easily extended to handle compressed models or those using low bit-width precision, making it a practical solution for efficient LLM research and prototype development.

\begin{table}[t!]
\caption{Comparison between our ELANA and the Zeus profiling framework~\citep{you2023zeus}.}
\centering
\begin{tabular}{lcccc}
\toprule
Framework        & {\bf ELANA (Ours)}  & Zeus (\texttt{ZeusMonitor}) \\\midrule
Usage & \makecell{Run a command from the terminal \\(\texttt{elana...}) without modifying the code} & \makecell{Insert \texttt{begin\_window}, \texttt{end\_window} \\calls around code blocks} \\\cline{1-3}
 Output & \makecell{Fine-grained latency and energy profiling \\results, including kernel-level execution\\ timeline (visualizable in \texttt{Perfetto})} & \makecell{High-level metrics such as total energy and time}\\\cline{1-3}
 Hardware & \makecell{Focused: NVIDIA GPUs for server \\and edge devices (e.g. Jetson series)}& \makecell{Broader: NVIDIA GPUs, AMD GPUs, \\CPUs, and Apple Silicon}\\\cline{1-3}
 Best For & \makecell{Profiling energy/latency of standardized LLM \\inference workflows at multiple granularities \\ and uncovering efficiency bottlenecks} & \makecell{Profiling energy and latency for\\ custom workflows}\\
\bottomrule
\end{tabular}
\label{Table: }
\end{table}

\section{Major Features and Profiling Results}

We introduce the major features and their profiling results of ELANA in this section.
To demonstrate the usage of ELANA, we profile several models, including Llama-3.1-8B \citep{meta2024llama}, Qwen-2.5-7B \citep{hui2024qwen2}, and the hybrid model Nemotron-H-8B \citep{blakeman2025nemotron}, on both A6000 GPUs (cloud) and Jetson AGX Thor 128GB and Orin Nano 8GB devices (edge).
For Orin Nano, we profile small language models, such as Llama-3.2-1B, Qwen2.5-1.5B.

\subsection{Hugging Face interface}

ELANA is designed to create the model to be profiled using the popular Hugging Face interface, as shown in the below code block.
Therefore, ELANA is able to support and profile models that are released on Hugging Face.
\begin{lstlisting}[language=Python]
def _build_model_and_tokenizer(self):
    tokenizer = AutoTokenizer.from_pretrained("model_hf_repo")
    model = AutoModelForCausalLM.from_pretrained("model_hf_repo")
\end{lstlisting}
This design choice allows researchers to integrate ELANA with emerging model architectures and newly developed compression algorithms with only a few lines of modification. For example,
\begin{lstlisting}[language=Python]
def _build_model_and_tokenizer(self):
    tokenizer = MyLocalTokenizer.from_pretrained("local_model_path") # local tokenizer
    model = MyLocalModel.from_pretrained("local_model_path") # local model
\end{lstlisting}
In summary, we aim to provide a simple and unified tool for future researchers to evaluate their new architectures and algorithms, without developing the profiling tool on their own.

\subsection{Model size profiling}
We use the SI (base-10) definition adopted by most storage manufacturers (i.e., 1\,GB = $1000^3$ bytes) as the default unit for profiling and reporting model size and cache size.
ELANA also provides the binary unit (GiB, where 1\,\text{GiB} = $1024^3$ bytes), commonly used in Linux and other operating systems, as an optional reporting memory unit.

\paragraph{Parameter and buffer size.}
ELANA reports the total parameter size of a model, including both trainable and non-trainable weights, to help users understand the memory footprint of the deployed model. 
In addition to parameters, ELANA also profiles auxiliary buffers such as positional embeddings and quantized layers. 
This enables practitioners to compare different compression algorithms and identify components that contribute most to memory usage, which is particularly important when deploying LLMs on memory-constrained edge devices.

\paragraph{KV and SSM cache size.}
During autoregressive generation, LLMs maintain intermediate states such as key–value (KV) caches in Transformers or recurrent state caches in State Space models (SSMs). 
These caches often dominate memory consumption, especially for long sequence generation or multi-request (\ie large batch size) serving. 
ELANA provides estimations of KV cache size for attention-based models and state-cache size for SSM-based architectures, allowing users to assess memory requirements under different serving workloads.

In Table~\ref{Table: Model and cache size profiling}, we profile the parameter size and KV cache size under different workloads of Llama-3.1-8B, Qwen-2.5-7B, and Nemotron-H-8B.
The profiling results are reported in GB.


\begin{table}[h!]
\caption{\textbf{(Model and cache size profiling.)}}
\centering
\begin{tabular}{@{}l|c|ccc@{}}
\toprule
\multirow{2}{*}[-0.5ex]{Model} & \multirow{2}{*}[-0.5ex]{\makecell{Param.\\size}}  & \multicolumn{3}{c}{Cache size} \\ 
\cmidrule(lr){3-5}
& & \makecell{bsize=1, L=1024} 
  & \makecell{bsize=128, L=1024} 
  & \makecell{bsize=128, L=2048} \\ 
\midrule
Llama-3.1-8B  & 16.06 GB & 0.13 GB & 17.18 GB & 34.36 GB \\
Qwen-2.5-7B   & 15.23 GB & 0.06 GB & 7.52 GB  & 15.03 GB \\
Nemotron-H-8B & 16.20 GB & 0.05 GB & 3.32 GB  & 6.64 GB  \\
\bottomrule
\end{tabular}
\label{Table: Model and cache size profiling}
\end{table}

\subsection{Latency profiling}

\paragraph{Time-to-first-token (TTFT, prefilling).}
TTFT measures the latency of the prefilling (\ie prompting) stage, where the entire input prompt is processed before the model generates the first output token. 
This metric reflects the latency of the initial forward pass, and is particularly important for interactive applications such as chat assistants or long-context summarization. 
ELANA provides accurate TTFT measurements by isolating the prefilling stage and reporting both raw latency and averaged statistics over multiple runs.
We prefill the model with random input prompts and profile the latency of TTFT.
Since input prompt lengths vary in real applications, we do \emph{not} cache CUDA graphs for the prefilling stage of the model inference.

\paragraph{Time-per-output-token (TPOT, generation).}
TPOT captures the average decoding latency per generated token during autoregressive generation (\ie decoding). 
Because the decoding stage is inherently sequential, TPOT serves as a key indicator of model efficiency for continuous token generation.
ELANA computes TPOT by recording the inter-token generation intervals and averaging them across the output sequence.
Before profiling TPOT, we prefill the KV cache using randomly generated inputs with the user-specified prompt length.
To maximize the throughput of the generation, we follow TensorRT-LLM \citep{nvidia2023tensorrtllm} and SGLang \citep{zheng2024sglang} to cache the CUDA graphs for the generation.

\paragraph{Time-to-last-token (TTLT, end-to-end).}
TTLT measures the complete end-to-end latency of the inference process, \ie from receiving the input prompt to generating the final output token.
This metric combines both prefilling and decoding latencies, providing a holistic view of the runtime of processing the requests.
ELANA reports TTLT alongside the decomposition into TTFT and TPOT, enabling practitioners to analyze bottlenecks and understand how different optimizations affect the overall inference experience.
We profile the TTLT using random input prompts and measure the end-to-end latency of processing a batch of requests under varying prompt and generation lengths.

In Table~\ref{Table: Latency and energy profiling on A6000}, we report the average latency over 100 runs on A6000 GPUs for all workloads in millisecond (ms), except for TTLT, for which we report the average over 20 runs.
We test ELANA on Jetson AGX Thor and Orin Nano, both edge GPU with 128 GB and 8GB unified memory, respectively.
The results of different workloads on Jetson series GPUs are reported in Table \ref{Table: Latency and energy profiling on Jetson Thor}.

\begin{table}[t!]
\caption{\textbf{(Latency and energy profiling on A6000.)} We profile various workloads on A6000 GPUs. Number of prefilling and generation tokens denote as $L=T_p+T_g$. The latency and energy are reported in millisecond (ms) and Joule (J), repectively.}
\centering
\begin{tabular}{@{}l|cc|cc|cc@{}}
\toprule
Model         & TTFT & J/Prom. & TPOT & J/Tok. & TTLT & J/Req. \\ \toprule
\multicolumn{7}{c}{nGPU=1, bsize=1, L=512+512} \\ \midrule
Llama-3.1-8B  & 94.30& 25.91    & 24.84& 6.80    & 12859.85& 3533.09 \\
Qwen-2.5-7B   & 88.41& 24.29    & 23.15& 6.44    & 12073.26& 3343.91 \\
Nemotron-H-8B & 87.72& 24.00    & 24.33& 6.67    & 12593.76& 3437.56 \\ \midrule
\multicolumn{7}{c}{nGPU=4, bsize=64, L=512+512} \\ \midrule
Llama-3.1-8B  & 1325.05&  476.50& 31.29& 10.94   &17329.35& 6131.45   \\
Qwen-2.5-7B   & 1192.98&  248.89& 26.48& 7.73    &14823.56& 5255.14  \\
Nemotron-H-8B & 1337.83&  478.82& 39.33& 13.86   &21300.36& 7499.34   \\ \midrule
\multicolumn{7}{c}{nGPU=4, bsize=64, L=1024+1024} \\ \midrule
Llama-3.1-8B  & 2788.39& 1044.31& 36.16& 12.72   &39935.79& 14219.00 \\
Qwen-2.5-7B   & 2454.50&  887.11& 28.66& 10.03   &32031.05& 11432.51  \\
Nemotron-H-8B & 2752.54& 1007.14& 39.40& 13.94   &42658.35& 15001.54  \\ \bottomrule
\end{tabular}
\label{Table: Latency and energy profiling on A6000}
\end{table}

\begin{table}[t!]
\caption{\textbf{(Latency and energy profiling on Jetson series.)} We profile various workloads on Jetson AGX Thor 128G and Orin Nano 8G. Number of prefilling and generation tokens denote as $L=T_p+T_g$. The latency and energy are reported in millisecond (ms) and Joule (J), repectively.}
\centering
\begin{tabular}{@{}l|cc|cc|cc@{}}
\toprule
Model         & TTFT & J/Prom. & TPOT & J/Tok. & TTLT & J/Req. \\ \toprule
\multicolumn{7}{c}{\textbf{Orin Nano 8GB} bsize=1, L=256+256} \\ \midrule
Llama-3.2-1B  &142.92& 0.42& 48.73& 0.06& 11601.61& 47.30  \\
Qwen2.5-1.5B  &249.89& 0.80& 60.66& 0.08& 14930.47& 60.21 \\ \toprule
\multicolumn{7}{c}{\textbf{Orin Nano 8GB} bsize=1, L=512+512} \\ \midrule
Llama-3.2-1B  &278.0& 1.12& 48.69& 0.06& 23590.22& 98.61  \\
Qwen2.5-1.5B  &359.30& 1.53& 61.43& 0.08& 30177.97& 123.94 \\ \toprule
\multicolumn{7}{c}{\textbf{AGX Thor 128GB} bsize=1, L=512+512} \\ \midrule
Llama-3.1-8B  & 147.49& 7.40& 97.60&  1.27& 32105.50& 633.19 \\
Qwen-2.5-7B   & 115.27& 6.39& 61.22&  0.88& 30875.60& 610.49 \\
Nemotron-H-8B & 147.29& 7.08& 101.73& 1.29& 33671.79& 655.17 \\ \midrule
\multicolumn{7}{c}{\textbf{AGX Thor 128GB} bsize=16, L=512+512} \\ \midrule
Llama-3.1-8B  & 2154.89& 140.83& 115.51& 1.87& 42317.18& 1176.06 \\
Qwen-2.5-7B   & 1879.78& 127.62& 109.18& 1.63& 35599.98& 930.34 \\
Nemotron-H-8B & 2008.94& 127.15& 140.08& 2.26& 53096.56& 1287.82\\ \midrule
\multicolumn{7}{c}{\textbf{AGX Thor 128GB} bsize=16, L=1024+1024} \\ \midrule
Llama-3.1-8B  & 4611.26& 296.29& 128.50& 2.37& 100605.99& 3041.79 \\
Qwen-2.5-7B   & 3848.15& 261.63& 117.19& 1.84& 78470.34&  2168.19\\
Nemotron-H-8B & 4388.04& 266.26& 141.01& 2.35& 104250.55& 2617.65 \\ \bottomrule
\end{tabular}
\label{Table: Latency and energy profiling on Jetson Thor}
\end{table}

\subsection{Energy profiling}
For energy profiling, we use \texttt{pynvml} to query the instantaneous power draw of the target GPU through NVIDIA’s NVML interface. 
On Jetson devices, we obtain the GPU power on the SoC using \texttt{jtop} from the \texttt{jetson-stats} toolkit, which reads the on-board power sensors. 
We sample the power usage every $0.1$ second and log all measurements. 
During latency profiling, a separate process runs concurrently to collect power readings, and we compute the average power over the corresponding measurement window. 
We combine this average power with the measured latency to obtain the energy consumption. 
We report Joules-per-prompt (J/Prompt) for TTFT, Joules-per-token (J/Token) for TPOT, and Joules-per-request (J/Request) for TTLT. 
In multi-GPU settings, we sum the average power across all participating GPUs when computing the final energy metrics.
In Table~\ref{Table: Latency and energy profiling on A6000}, we report the average energy over 100 runs on A6000 GPUs for all workloads in Joule, except for J/Req., for which we report the average over 20 runs.
We also report the energy costs for the experimental models on Jetson AGX Thor and Orin Nano in Table \ref{Table: Latency and energy profiling on Jetson Thor}.

\subsection{Fine-grained kernel profiling}

\begin{figure*}
    \centering
    \includegraphics[width=\linewidth]{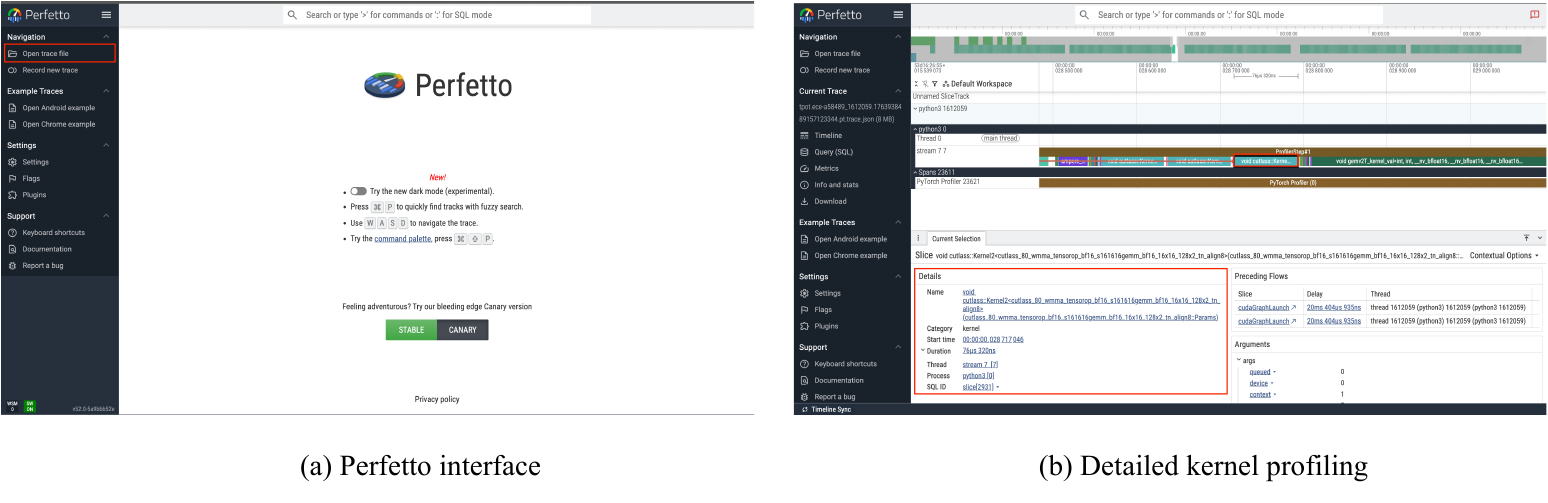}
    \caption{\textbf{(Detailed kernel profiling results on Perfetto.)}}
    \label{fig: Detailed kernel profiling results on Perfetto}
\end{figure*}

\paragraph{PyTorch Profiler.}
To support fine-grained analysis, ELANA optionally integrates the PyTorch Profiler\footnote{PyTorch Profiler – the new and improved performance tool:  \url{https://docs.pytorch.org/docs/stable/profiler.html}} and Holistic Trace Analysis (HTA) \footnote{Holistic Trace Analysis: \url{https://github.com/facebookresearch/HolisticTraceAnalysis}}  for capturing low-level execution traces, operator runtimes, and kernel-level statistics.
This feature enables users to diagnose performance bottlenecks, analyze GPU utilization, and inspect the impact of kernel fusion, quantization, or model compression techniques.
The profiling results can be exported for visualization with tools such as Perfetto\footnote{Perfetto - System profiling, app tracing and trace analysis:  \url{https://ui.perfetto.dev/}} for further analysis, providing a detailed view of runtime behavior beyond high-level latency metrics.
Figure \ref{fig: Detailed kernel profiling results on Perfetto} shows an example of the profiling results on Perfetto.

\section{Conclusion}
ELANA provides a lightweight, accessible, and extensible framework for evaluating the efficiency of large language models across cloud and edge GPUs. By offering comprehensive profiling of model size, KV cache memory, latency characteristics, and optional energy consumption, ELANA enables both practitioners and researchers to gain deeper insights into the performance bottlenecks of modern LLMs. Its compatibility with Hugging Face APIs and adaptability for newly developed models or compression algorithms further position ELANA as a practical tool for advancing research on efficient model design and deployment. We hope ELANA serves as a foundation for reproducible benchmarking, facilitates fair comparisons across models and systems, and accelerates the development of next-generation, resource-efficient LLMs.






\clearpage

%

\section*{Acknowledgments}
This work was supported in part by the ONR Minerva program, NSF CCF Grant No. 2107085, iMAGiNE - the Intelligent Machine Engineering Consortium at UT Austin, UT Cockrell School of Engineering Doctoral Fellowships.

\bibliographystyle{plainnat}   
\bibliography{biblio}          




\end{document}